\documentclass[letter]{ptptex}
\usepackage[dvips]{graphicx,color}
\usepackage{multirow}
\usepackage{bm}
\usepackage{wrapft}

\def\fun#1#2{\lower3.6pt\vbox{\baselineskip0pt\lineskip.9pt
\ialign{$\mathsurround=0pt#1\hfil##\hfil$\crcr#2\crcr\sim\crcr}}}

\def\a{\alpha}

\def\d{\delta}
\def\ve{\varepsilon}

\newcommand{\beq}{\begin{equation}}
\newcommand{\eeq}{\end{equation}}
\newcommand{\bea}{\begin{eqnarray}}
\newcommand{\eea}{\end{eqnarray}}

\newcommand{\m}{\mu}

\newcommand{\bfi}[1]{\mbox{\boldmath $#1$}}
\newcommand{\bfis}[1]{\mbox{\boldmath ${\scriptstyle #1}$}}

\newcommand{\vk}{{\bfi k}}

\newcommand{\vr}{{\bfi r}}

\newcommand{\vir}{{\bfis r}}

\def\non{\nonumber}

\title{
A Smoothing Method of Discrete Breakup $S$-matrix Elements
in the Theory of Continuum-Discretized Coupled Channels
}

\author{
Takuma \textsc{Matsumoto}$^1$, Tomoaki \textsc{Egami}$^2$,
Kazuyuki \textsc{Ogata}$^2$, Masanobu \textsc{Yahiro}$^2$%
}

\inst{
$^1$RIKEN Nishina Center, Hirosawa 2-1, Wako, Saitama 351-0198, Japan\\
$^2$Department of Physics, Kyushu University, Fukuoka 812-8581, Japan
}

\abst{
 We present a practical way of smoothing discrete breakup $S$-matrix elements
 calculated by the continuum-discretized coupled-channel method
 (CDCC). This method makes the smoothing procedure much easier.
 The reliability of the smoothing method is
 confirmed for the three-body breakup reactions,
 $^{58}$Ni($d,pn$) at 80~MeV and
 $^{12}$C($^6$He,$^4$He$^2n$) at 229.8~MeV.
}

\begin{document}
\maketitle

\section{Introduction}
\label{sec:intro}

Secondary beam experiments are opening up new physics on
unstable nuclei away from the stability line in the nuclear chart.
Such unstable nuclei have exotic properties, e.g., the halo
structure.~\cite{Tanihata1,Tanihata2,Hansen} \
In scattering of such a weakly bound projectile,
it easily breaks up into its constituents, and hence the
reaction should be described as at least a three-body
scattering.

One of the most reliable methods for treating the projectile
breakup processes in the wide
range of incident energy is the method of continuum-discretized coupled
channels (CDCC).~\cite{CDCC-review1,CDCC-review2} \ In CDCC, the
scattering wave function of the total system is expanded with
the complete set of bound and continuum states of the projectile.
The continuum states are
classified by linear and angular momenta, $k$ and $\ell$, and
truncated at upper bounds, $k_{\rm max}$ and $\ell_{\rm max}$, respectively.
This truncation is based on the assumption that
the breakup processes take place in the truncated space.
The $k$-continuum is divided into small bins and the
continuum states in each bin are averaged into a single state. This
discretization procedure is called the average discretization method.
The space spanned by the bound states and the finite number of
discretized continuum states are called the modelspace.
The $S$-matrix elements calculated with CDCC converge as the modelspace
is extended.~\cite{CDCC-convergence-1,CDCC-convergence-2} \
The converged CDCC solution is the unperturbed solution of
the distorted Faddeev equations, and corrections to the solution are
negligible within the spatial region in which the breakup processes take
place.~\cite{CDCC-foundation1,CDCC-foundation2}

The average discretiztation method has widely been
used so far, but its application is limited in
the three-body breakup processes in scattering of two-body
projectiles. The average method is not feasible for the four-body
breakup processes in scattering of three-body projectiles,
since it requires exact three-body continuum
states of the projectile that are quite difficult to obtain.
This problem can be solved
by using the pseudostate discretization
method~\cite{Matsumoto, Egami, Matsumoto2,Matsumoto3, Matsumoto4, Moro1,
THO-CDCC} in which
the continuum states $\{\psi(\bm{k})\}$
are replaced by pseudostates $\{\hat{\Phi}_{i}\}$
obtained by diagonalizing the internal Hamiltonian of the projectile in
a space spanned by $L^2$-type basis functions.
One can adopt the
Gaussian~\cite{Matsumoto,Egami} or the transformed harmonic oscillator
(THO)~\cite{Moro1} basis as the $L^2$-type basis functions.
The validity of the pseudostate discretization method was confirmed in
scattering of two-body projectile by the good agreement
between the CDCC solutions obtained by the pseudostate discretization and
the average methods.~\cite{Matsumoto, Egami, Matsumoto2,Matsumoto3, Matsumoto4}

For four-body reaction systems, CDCC calculations with
the pseudostate discretization method based on
Gaussian~\cite{Matsumoto2, Matsumoto3,Matsumoto4} or
THO~\cite{THO-CDCC} basis functions were successful in describing
the elastic scattering at not only high
energies but also low energies near the Coulomb barrier.
Thus, the back-coupling effects of four-body breakup processes
on the elastic scattering, i.e., {\it virtual} four-body breakup
processes, are well described by CDCC based on
the pseudostate discretization method.

The $S$-matrix elements calculated with CDCC, $\hat{S}_{i} $,
are discrete in $\vk$,
while 
the exact ones $S(\bm{k})$ are continuous.
Thus, one needs a way of smoothing $\hat{S}_{i}$ in order to analyze
real breakup processes, i.e. breakup reactions themselves.
In principle, this can be possible by calculating
the overlap
$\mathcal{F}_{i}(\bm{k})=\langle \psi(\bm{k})|\hat{\Phi}_{i}\rangle$
between the exact continuum states $\psi(\bm{k})$ and
the pseudostates $\hat{\Phi}_{i}$; see \S \ref{sub:smoothing} for the detail.
In practice, this smoothing factors $\mathcal{F}_{i}(\bm{k})$
can easily be obtained in three-body
breakup reactions of two-body projectiles, but it is quite hard in four-body
ones of three-body projectiles. Very recently, we have proposed a smoothing
method applicable to four-body breakup reactions,~\cite{Egami2008} \
but it still requires heavy numerical calculations and hence
not so practical. Thus, it is highly expected that
an accurate and practical method of smoothing $\hat{S}_{i}$ is proposed.

The complex scaling method (CSM)~\cite{ABC} is a powerful tool of
solving many-body resonance and weakly bound states.
Recently, CSM was extended so that it can treat
the electromagnetic transition of the core+nucleon+nucleon system such as
$^{6}$He and $^{11}$Li
from its ground state to the continuum ones.~\cite{CSM6He,CSM11Li} \
As a merit of CSM, the calculated transition strength is obtained
as a continuum spectrum. This implies that an accurate and practical method
of smoothing $\hat{S}_{i}$ can be constructed with CSM.

In this Letter, we present a simple formula that makes it possible
to smooth $\hat{S}_{i}$ accurately, using CSM.
This formula is useful for not only three-body but also four-body
breakup reactions.
Validity of the formula is tested in the three-body breakup
reactions, $^{58}$Ni($d,pn$) at 80~MeV and
$^{12}$C($^6$He,$^4$He$^2n$) at 229.8~MeV,
since in such three-body reactions
the {\lq\lq}exact'' $S$-matrix elements are obtainable
by calculating $\mathcal{F}_{i}(\bm{k})$ exactly.
This test is inevitable to proceed to CDCC calculations for
four-body breakup processes in future.

This Letter is constructed as follows. In Sec.~\ref{sec:formulation}, we first
recapitulate CDCC and the pseudostate discretization method, and
present an accurate and practical smoothing formula of $\hat{S}_{i}$.
Test calculations are done in Sec.~\ref{sec:test}.
Section~\ref{sec:summary} gives a summary.

\section{Formulation}
\label{sec:formulation}

\subsection{CDCC and the pseudostate discretization}
\label{sub:cdcc}

For simplicity, we consider the three-body (A+b+c) system,
where particles A, b and c are assumed to be structureless.
It is straightforward to extend the present formulation to the four-body
system in which the projectile consists of three constituents.
The total Hamiltonian of the A+b+c system is
\begin{align}
 H &=T_R+U+H_{\rm B},\\
 U &=U_b(\bm{R}_b)
    +U_c(\bm{R}_c)+V^{\rm Coul}(\bm{R}),\\
 H_{\rm B}&=T_r+V_{bc},
\end{align}
where $\bm{R}$ is the relative coordinate between the center-of-mass
(c.m.) of the projectile B=b+c and A.
Coordinate $\bm{R}_{\rm x}$ (x $=$ b and c) denotes the relative
coordinate between x and A, and $T_{\xi}$ ($\xi=\bm{R}$ and $\bm{r}$)
is the kinetic-energy operator associated with $\xi$.
Interaction $V_{\rm bc}$ is the potential
between b and c, and
$U_{\rm x}$ is the nuclear part of the optical potential
between x and A.
Meanwhile the Coulomb interaction $V^{\rm Coul}$ between
B and A is treated
approximately as a function of $\bm{R}$ only, i.e., we neglect Coulomb
breakup processes.

In the pseudostate discretization
method,~\cite{Matsumoto, Egami, Matsumoto2,Matsumoto3, Matsumoto4} \
the pseudostates \{$\Hat{\Phi}_{i}$\} are
obtained by diagonalizing the internal Hamiltonian $H_{\rm B}$
in a space spanned by the $L^2$-type Gaussian basis
functions~\cite{H-Ka-Ki}:
\begin{align}
\langle \hat{\Phi}_{i}|H_{\rm B}| \hat{\Phi}_{i'} \rangle =
\delta_{ii'}\hat{\ve}_{i},
\end{align}
where $\hat{\ve}_{i}$ is the eigenenergy of $\hat{\Phi}_{i}.$
The subscript $i$ denotes a set of quantum numbers, i.e.
the energy index $n$ of the pseudostates,
the angular momentum $\ell$ between b and c,
and its projection on the $z$-axis $m$.
The eigenstates of $H_{\rm B}$ with negative and positive energies
correspond to the bound state(s) and the pseudostates, respectively,
and the latter are regarded as discretized continuum states.

The basic assumption of CDCC is that
the breakup processes take place in a model
space~\cite{Matsumoto3,Matsumoto4}
\begin{align}
 {\cal P}&=\sum_{i}|\hat{\Phi}_{i}\rangle \langle\hat{\Phi}_{i}|.
 \label{eq:com-set}
\end{align}
Validity of this assumption is justified by the fact that
the calculated elastic and total breakup
cross sections of the three- and four-body scatterings
converge as the modelspace is
extended.~\cite{Matsumoto, Egami, Matsumoto2,Matsumoto3, Matsumoto4} \
Thus, one may regard $\{\hat{\Phi}_i\}$ as a complete set
in describing the reaction processes considered. We henceforth
call $\{\hat{\Phi}_i\}$ the approximate complete set in this
meaning.

We solve the three-body
Schr\"odinger equation in the modelspace ${\cal P}$,
\begin{align}
{\cal P}[H-E_{\rm tot}]{\cal P}|\Psi^{\rm CDCC} \rangle =0,
\label{eq:4b-Schr}
\end{align}
where $E_{\rm tot}$ is the total energy of the system.
The total wave function $\Psi^{\rm CDCC}$
is expanded by the approximate
complete set $\{\hat{\Phi}_{i}\}$:
\begin{align}
 |\Psi^{\rm CDCC} \rangle
 = \sum_{i} | \hat{\Phi}_{i}, \hat{\chi}_{i}\rangle,
 \label{eq:CDCC-1}
\end{align}
where $|\hat{\Phi}_{i},\hat{\chi}_{i}\rangle
=|\hat{\Phi}_{i}\rangle\otimes|\hat{\chi}_{i}\rangle$
and $i=0$ denotes the elastic channel and others ($i\ne 0$)
the breakup channels.
The expansion coefficient $|\hat{\chi}_{i}\rangle$ describes
the relative motion between B (in state $\hat{\Phi}_{i}$)
and A.
The intrinsic energy $\hat{\ve}_{i}$ of B
and the relative momentum $\hat{P}_{i}$ between B and A
satisfy the energy conservation,
$E_i \equiv \hbar^2\hat{P}^2_i/(2\mu)
 =E_{\rm tot}-\hat{\ve}_{i}$, with the reduced mass $\mu$ between B and A.

Multiplying Eq.~\eqref{eq:4b-Schr} by
$\langle\hat{\Phi}_{i}|$ from the left, we can obtain a set of coupled
differential equations for $|\hat{\chi}_{i}\rangle$, called CDCC
equation,
\begin{align}
 [T_R+\hat{U}_{i,i}-E_i]|\hat{\chi}_{i}\rangle
 = -\sum_{i'\ne i} \hat{U}_{i,i'}|\hat{\chi}_{i'}\rangle.
 \label{eq:CDCC-eq}
\end{align}
The coupling potential $\hat{U}_{i,i'}$ is defined by
\begin{align}
 \hat{U}_{i,i'}
 &=\langle \hat{\Phi}_{i} | U | \hat{\Phi}_{i'}\rangle.
 \label{eq:4B-CP}
\end{align}
The CDCC equation \eqref{eq:CDCC-eq} is solved under the usual
boundary condition for $\langle \bm{R}|\hat{\chi}_i\rangle\equiv\hat{\chi}_i
(\bm{R})$.~\cite{CDCC-review1,CDCC-review2}

\subsection{Smoothing method}
\label{sub:smoothing}

In this subsection, we present a new way of smoothing
discrete breakup $S$-matrix elements $\hat{S}_{i}$ obtained by CDCC.
The exact breakup $T$-matrix element to the exact continuum state
$\psi(\bm{k})$ of B is given by
\begin{align}
 T^{\rm EX}(\bm{k},\bm{P})
 &=\langle \psi(\bm{k}), P|U|\Psi\rangle \
 \label{eq:t-mat}
\end{align}
with $ |\psi(\bm{k}), P\rangle =
|\psi(\bm{k})\rangle\otimes|P\rangle$, where $\bm{k}$
and $\bm{P}$ are momenta in the asymptotic region associated
with the coordinates $\bm{r}$ and $\bm{R}$, respectively;
$|\psi(\bm{k})\rangle$ is the exact two-body wave function of
B with energy $\ve$ satisfying
\begin{align}
[H_{\rm B}-\ve]|\psi(\bm{k})\rangle=0,
\label{eq:Sch-B}
\end{align}
and $|P\rangle$ is the plane wave function satisfying
\begin{align}
  [T_R-(E_{\rm tot}-\ve)]|P\rangle =0.
\end{align}
The exact three-body (A+b+c) wave function $\Psi$
can be replaced by the corresponding CDCC wave function
$\Psi^{\rm CDCC}$ with good accuracy.
Inserting the approximate complete set
${\cal P}$ of Eq.~\eqref{eq:com-set}
between the bra-vector and the operator $U$ of the
right hand side of Eq.~\eqref{eq:t-mat}, we can obtain the
approximate smooth $T$-matrix elements $T(\bm{k}, \bm{P})$:
\begin{align}
 T(\bm{k}, \bm{P})
 &=
 \sum_{i}\langle \psi(\bm{k})|\hat{\Phi}_{i}\rangle
 \langle\hat{\Phi}_{i}, \hat{P}_i | U |\Psi^{\rm CDCC} \rangle\notag\\
 &\equiv \sum_{i}\mathcal{F}_{i}(\bm{k})  \hat{T}_{i},
\label{eq:t-mat2}
\end{align}
where
$|\hat{\Phi}_{i}, \hat{P}_i \rangle=
|\hat{\Phi}_{i}\rangle\otimes|\hat{P}_i\rangle$,
$\mathcal{F}_{i}(\bm{k})$ is the smoothing factor defined by
\begin{align}
  \mathcal{F}_{i}(\bm{k})
 &=\langle \psi(\bm{k})|\hat{\Phi}_{i}\rangle,
\end{align}
and $\hat{T}_{i}$ is the breakup $T$-matrix element of CDCC
defined by
\begin{align}
 \hat{T}_{i}
 &=\langle\hat{\Phi}_{i}, \hat{P}_i | U | \Psi^{\rm CDCC}\rangle.
\end{align}
Since the breakup $T$-matrix elements are
proportional to the breakup $S$-matrix elements,
Eq.~\eqref{eq:t-mat2} is reduced to
\begin{align}
 S(\bm{k},\bm{P})
 = \sum_{i}\mathcal{F}_{i}(\bm{k}) \hat{S}_{i}.
 \label{eq:smooth-S}
\end{align}

The smoothing factor $\mathcal{F}_{i}(\bm{k})$
is the overlap between the pseudostate $\hat{\Phi}_{i}$ and
the exact continuum state $\psi(\bm{k})$.
We propose to use the complex-scaling method (CSM)~\cite{ABC}
for the calculation of $\mathcal{F}_{i}(\bm{k})$.
This new smoothing method is applicable
to not only the two-body projectile but also the three-body one,
as shown below.

In CSM, the scaling transformation operator $C(\theta)$ and
its inverse operator are defined by
\begin{eqnarray}
\langle \vr | C(\theta)|f \rangle =e^{i3/2\theta}f(\vr e^{i\theta}), \quad
\langle g | C^{-1}(\theta)| \vr \rangle
= \{ e^{-i3/2\theta}f(\vr e^{-i\theta}) \}^{*} ,
\end{eqnarray}
since
\bea
\langle g | C^{-1}(\theta) C(\theta)|f \rangle =
 \int d\vr \langle g | C^{-1}(\theta)| \vr \rangle
           \langle \vr | C(\theta)|f \rangle =
           \langle g | f \rangle
\eea
for any normalizable  states $g$ and $f$, and
hence $C^{-1}(\theta) C(\theta)=1$.
The transformed Scrh\"{o}dinger equation
is defined by
\begin{eqnarray}
 (H_{\rm B}^\theta-\ve^\theta)|\psi^\theta \rangle=0, \quad
 \langle {\tilde \psi}^\theta |(H_{\rm B}^\theta-\ve^\theta)=0
 \label{eq:Sch-B-th}
\end{eqnarray}
with
\begin{eqnarray}
\langle \vr | H_{\rm B}^\theta | \vr' \rangle =
\langle \vr |C(\theta)H_{\rm B}C^{-1}(\theta) | \vr' \rangle
 =\left[-e^{-2i\theta}\frac{\nabla_{\vir}^2}{2\m_{bc}}
 +V_{bc}(\vr e^{i\theta})\right] \d(\vr-\vr'),
 \quad
\end{eqnarray}
where $\m_{bc}$ is the reduced mass between b and c.
The transformed Hamiltonian $H^\theta$ is not hermitian
when $\theta \neq 0$ and thus the ket vectors $|\psi^\theta \rangle$ and
the bra ones $\langle {\tilde \psi}^\theta |$ are biorthogonal to each
other.~\cite{Berggren1} \
Properties of $|\psi^\theta \rangle$ and $\ve^\theta$ are summarized
as follows.

\begin{enumerate}
  \item
  In the case that
  $|\psi^\theta \rangle$ is a bound state $|\psi_{j}^\theta \rangle$
  with a negative energy
  $\ve_{j}^\theta=-\hbar^2(k_{j}^\theta)^2/(2\m_{bc})$,
  $\langle \vr|\psi_{j}^\theta \rangle$ has an asymptotic form
  $\exp(-k_{j}^\theta r e^{i\theta})$,
  so that it becomes a  normalizable $L^2$-type
  function for $ 0 \le \theta < \pi/2$.
  Operating $C(\theta)$ on Eq.~\eqref{eq:Sch-B} leads to
  $(H_{\rm B}^{\theta}-\ve)C(\theta)|\psi \rangle=0$.
  Comparing this equation with the first equation of
  Eq.~\eqref{eq:Sch-B-th},
  one can find that
  $|\psi_{j}^{\theta} \rangle=C(\theta)|\psi_{j} \rangle$
  and the eigenenergy $\ve_{j}^\theta$ is
  independent of $\theta$.

  \item
  In general a resonance state
  $|\psi_{\a}^\theta \rangle$ has an energy
  $\ve_{\a}^\theta=\hbar^2(k_{\a}^\theta e^{-i\phi})^2/(2\m_{bc})$ with
  positive $\phi$ and $k_{\a}^\theta$. The state has an asymptotic form
  $\exp(i k_{\a}^\theta r e^{i(\theta-\phi)})$, and hence
  it is a normalizable $L^2$-type function for $\theta >\phi$.
  Let us take the case that $\theta_1 > \theta > \phi$.
  Operating $C(\theta_1-\theta)$ on the first equation
  of Eq.~\eqref{eq:Sch-B-th} and using
  $C(\theta_1)=C(\theta_1-\theta)C(\theta)$, we obtain 
  $(H_{\rm B}^{\theta_1}-\ve_{\a}^{\theta})C(\theta_1-\theta)
  |\psi_{\a}^{\theta} \rangle=0$.
  Identifying this equation with
  $(H_{\rm B}^{\theta_1}-\ve_{\a}^{\theta_1})|\psi_{\a}^{\theta_1}
  \rangle=0$
  leads to     the fact that
  $|\psi_{\a}^{\theta_1} \rangle =C(\theta_1-\theta)
  |\psi_{\a}^{\theta} \rangle$ and
  $\ve_{\a}^{\theta}$ does not depend on $\theta$ if $\theta > \phi$.

  \item
  The scattering state $|\psi^\theta(\vk) \rangle$ should be
  normalizable as $\langle \psi^\theta(\vk)|\psi^\theta(\vk')
  \rangle=\d(\vk-\vk')$. This means that the incident-wave part of the state
  is $\varphi_0=(2\pi)^{-3/2}\exp(i\vk \cdot \vr)$ and hence $\vk$ has to be
  transformed by the scaling transformation as $\vk \to \vk e^{-i\theta}$.
  Therefore, the scattering state has an energy
  $\ve^\theta(\vk)=(\hbar\vk e^{-i\theta})^2/(2\m_{bc})$, so that
  $|\psi^\theta(\vk) \rangle \neq C(\theta) | \psi^{\theta=0} (\vk)
  \rangle$. Note that $\langle \vr | C(\theta) | \psi^{\theta=0} (\vk)
  \rangle$ diverges at large $r$ and thus not normalizable.
\end{enumerate}

The scattering state $|\psi (\vk) \rangle$ at $\theta=0$
satisfies the integral equation
\begin{eqnarray}
 |\psi(\vk) \rangle &=& |\varphi_0\rangle
  + \frac{1}{\ve-H_{\rm B}} V_{bc} |\varphi_0 \rangle
  =|\varphi_0\rangle
  +C^{-1}(\theta)
  \frac{1}{\ve-H_{\rm B}^\theta}
  V_{bc}^\theta C(\theta) |\varphi_0 \rangle ,
\label{eq:integral}
\end{eqnarray}
where
use has been made of $C^{-1}(\theta)C(\theta)=1$ in
the second equality of Eq.~\eqref{eq:integral}
and $|\varphi_0\rangle$ describes the plane wave satisfying
\begin{eqnarray}
 \left[T_r-\ve\right]|\varphi_0\rangle=0.
\end{eqnarray}
One can thus obtain the smoothing factor by
\begin{eqnarray}
\langle {\hat \Phi}_i |\psi\rangle &=&\langle {\hat \Phi}_i|\varphi_0\rangle
  +\langle {\hat \Phi}_i|C^{-1}(\theta)
  \frac{1}{\ve-H_{\rm B}^\theta}
  V_{\rm bc}^\theta C(\theta) |\varphi_0 \rangle ,
  \label{LS_tr}
\end{eqnarray}
if the full propagator
$G^\theta=(\ve-H_{\rm B}^\theta)^{-1}$
is given for positive $\ve$.

The eigenstates
$\{\psi_{j}^\theta, \psi_{\a}^\theta, \psi^\theta(\vk) \}$ are
normalizable and can form the complete set.~\cite{Berggren2} \
Thus, the spectral representation of $G^\theta$ is given by
\beq
G^\theta= \sum_{j} |\psi_{j}^\theta \rangle \frac{1}{\ve - \ve_j^\theta}
\langle {\tilde \psi}_j^\theta | +
\sum_{\a} |\psi_{\a}^\theta \rangle \frac{1}{\ve - \ve_{\a}^\theta}
\langle {\tilde \psi}_{\a}^\theta | +
\int d\vk |\psi^{\theta} (\vk) \rangle \frac{1}{\ve - \ve^{\theta}(\vk)}
\langle {\tilde \psi}^{\theta} (\vk) |.
\label{exact-spectral-rep}
\eeq
In this representation, the resonance part (the second term
of the right-hand side) is separated out from the scattering part
(the third term). This is useful in investigating a role of
the resonance channel on the elastic scattering and the breakup reactions.

The exact representation Eq.~\eqref{exact-spectral-rep}
is still complicated and
not so useful in particular for systems more complicated
than the two-body system.
The full propagator $\langle \vr |G^{\theta} |\vr'\rangle
(=G^{\theta}(\vr,\vr'))$ satisfies the equation
\bea
G^{\theta}(\vr,\vr')
&=&G_0^\theta(\vr,\vr')+ \int d\vr''
G_0^\theta(\vr,\vr'') V(e^{i\theta}\vr'')
G^\theta(\vr'',\vr') \non \\
&=&G_0^\theta(\vr,\vr')+
\int d\vr''
G_0^\theta(\vr,\vr'') V(e^{i\theta}\vr'')
G_0^\theta(\vr'',\vr') + \cdots ,
\label{full-propagator-eq}
\eea
where
\beq
G_0^\theta(\vr,\vr')
=\langle \vr | \frac{1}{\ve - e^{-2i\theta}T_{r}}
| \vr' \rangle =
-\frac{\m_{bc}e^{2i\theta}}{2\pi \hbar^2}
\frac{\exp[i \vk\cdot|\vr-\vr'| e^{i\theta}]}{|\vr-\vr'|}.
\eeq
The free propagator $G_0^\theta(\vr,\vr')$ is
an exponentially damping function of $|\vr-\vr'|$ when
$0< \theta \le \pi$. Each term in the last form
of Eq.~\eqref{full-propagator-eq} has this property. Therefore,
the full propagator $G^{\theta}(\vr,\vr')$ keeps the damping property.
This assures that $G^{\theta}$ can be expressed by a superposition of
$L^2$-type basis functions; namely,
\beq
G^{\theta} \approx {\cal P}^{\theta} G^{\theta} {\cal P}^{\theta},
\eeq
where the operator ${\cal P}^{\theta}$ represents
the modelspace spanned by $L^2$-type basis functions.
The most convenient representation of ${\cal P}^{\theta}$ is
\beq
{\cal P}^{\theta}=\sum_i |{\it \Phi}^\theta_i\rangle
\langle\tilde{\it \Phi}_i^\theta| ,
\eeq
where the $\Phi^\theta_i$ are eigenstates of $H_{\rm B}^\theta$
obtained by diagonalizing it 
with $L^2$-type basis functions:
\begin{eqnarray}
\langle {\tilde \Phi}^\theta_i |H_{\rm B}^\theta| \Phi^\theta_{i'} \rangle
=\ve_i^\theta \delta_{ii'}.
  \label{theta-eigenstates}
\end{eqnarray}
One can find the approximate relation
$\langle {\tilde \Phi}^\theta_i |
G^\theta | \Phi^\theta_{i'} \rangle \approx (\ve - \ve_i^\theta)^{-1}
\d_{ii'}$,
because
\beq
\d_{ii'}=\langle {\tilde \Phi}^\theta_i | \Phi^\theta_{i'} \rangle
=\langle {\tilde \Phi}^\theta_i |(\ve-H_{\rm B}^\theta)G^\theta
| \Phi^\theta_{i'} \rangle \approx
\langle {\tilde \Phi}^\theta_i |(\ve-H_{\rm B}^\theta){\cal P}^\theta G^\theta
| \Phi^\theta_{i'} \rangle =
(\ve - \ve_i^\theta)\langle {\tilde \Phi}^\theta_i |
G^\theta | \Phi^\theta_{i'} \rangle.
\eeq
Using this approximate relation, we can obtain
\begin{eqnarray}
G^{\theta} \approx
  \sum_{ii'} |{\it \Phi}^\theta_i\rangle
  \langle\tilde{\it \Phi}_i^\theta| G^\theta
|{\it \Phi}_{i'}^\theta\rangle
  \langle\tilde{\it \Phi}_{i'}^\theta |
  =\sum_i |{\it \Phi}^\theta_i\rangle
  \frac{1}{\ve-\ve_i^\theta} \langle\tilde{\it \Phi}_i^\theta | .
  \label{Green's-function}
\end{eqnarray}
This approximation has been applied to calculations of
the continuum level density~\cite{SMK},
the scattering amplitude~\cite{KSK}, and
the strength function of the electromagnetic
transition~\cite{CSM6He,CSM11Li}.
Validity of the approximate expression Eq.~\eqref{Green's-function} in
the CDCC framework will be justified,
if the breakup $S$-matrix elements converge
as the modelspace ${\cal P}^\theta$ is extended.
Inserting Eq.~\eqref{Green's-function} into Eq.~\eqref{LS_tr},
we obtain the equation for the smoothing factor:
\begin{eqnarray}
\langle {\hat \Phi}_i |\psi\rangle \approx
\langle {\hat \Phi}_i|\varphi_0\rangle
  + \sum_{i'i''} \frac{1}{\ve-\ve_{i'}^\theta}
  \langle {\hat \Phi}_i|C^{-1}(\theta)
   |{\it \Phi}^\theta_{i'}\rangle
   \langle \tilde{\it \Phi}_{i'}^\theta |
  V^\theta |{\it \Phi}^\theta_{i''}\rangle
  \langle \tilde{\it \Phi}_{i''}^\theta | C(\theta) |\varphi_0 \rangle .
  \label{eq-for-smoothing}
\end{eqnarray}
When the Gaussian basis functions are taken, the matrix elements
$\langle {\hat \Phi}_i|C^{-1}(\theta) |{\it \Phi}^\theta_{i'}\rangle$ and
$\langle \tilde{\it \Phi}_{i''}^\theta | C(\theta) |\varphi_0 \rangle$
are analytically obtained.
The matrix elements $\langle \tilde{\it \Phi}_{i'}^\theta |
V^\theta |{\it \Phi}^\theta_{i''}\rangle$ are also easily obtained by
  making a single integral.
Thus, the final formula Eq.~\eqref{eq-for-smoothing}
for the smoothing factor is quite useful.

\section{Test calculations}
\label{sec:test}

In this section, we test the validity of the smoothing formula,
Eq. \eqref{eq-for-smoothing}, in two cases of
the $^{58}$Ni($d$,$pn$) reaction at 80 MeV and
the $^{12}$C($^6$He,$^4$He$^2n$) reaction at 229.8 MeV.
In the former, we have chosen $d$ as a typical projectile
with no resonance and treat it with the $n$+$p$ model.
In the latter, we have selected $^6$He as a projectile with resonance and
for simplicity will treat it with the two-body $^2n$+$^4$He model.
In principle, $^6$He should be treated as the three-body
$n$+$n$+$^4$He system. However, since our interest is
the $2^{+}$ resonance in the present test, we use
the dineutron ($^2n$+$^4$He)
model that can describe the $2^{+}$ resonance reasonably well.

The pseudostates ${\hat \Phi}_i$ in the CDCC calculations are constructed
by diagonalizing $H_{\rm B}$ with the complex-range Gaussian basis functions,
which are found to effectively describe breakup processes.~\cite{Matsumoto} \
The explicit forms of the basis functions are
\begin{eqnarray}
 \phi_{j\ell}^{\rm C}(r)&=&
  r^\ell\exp\left[-(r/a_j)^2\right]\cos\left[\pi/2(r/a_j)^2\right],
\\
 \phi_{j\ell}^{\rm S}(r)&=&
  r^\ell\exp\left[-(r/a_j)^2\right]\sin\left[\pi/2(r/a_j)^2\right] ,
\end{eqnarray}
where  $j=1$--$N$.
The range parameter $a_j$ is taken as a geometric progression,
\begin{eqnarray}
 a_j&=&a_1(a_N/a_1)^{(j-1)/(N-1)}.
\end{eqnarray}
The number of the basis functions is $2N$ for each $\ell$.

On the other hand, the complex scaled states ${\Phi}_i^\theta$
are constructed by diagonalizing $H_{\rm B}^\theta$
with the conventional real-range Gaussian functions,
\begin{eqnarray}
 \phi_{j\ell}(r)&=&r^\ell\exp[-(r/a_j)^2], \;(j=1\mbox{--}N),
\end{eqnarray}
which are valid for describing the
approximate expression of $G^\theta$ as shown in Ref.~\citen{KSK}.  

\begin{figure}[tb]
\begin{center}
 \includegraphics[clip,width=0.4\textwidth]{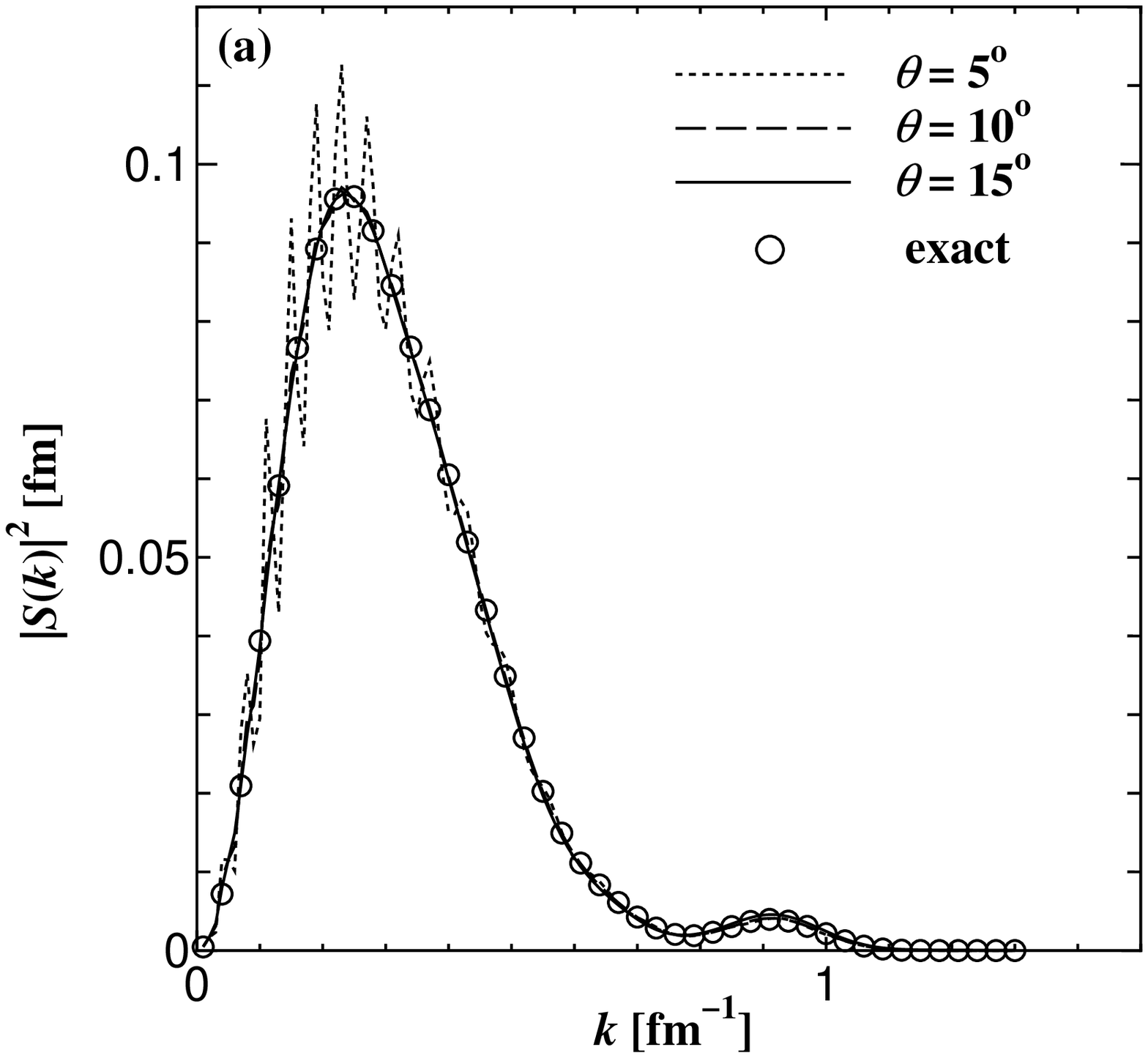}
 \includegraphics[clip,width=0.4\textwidth]{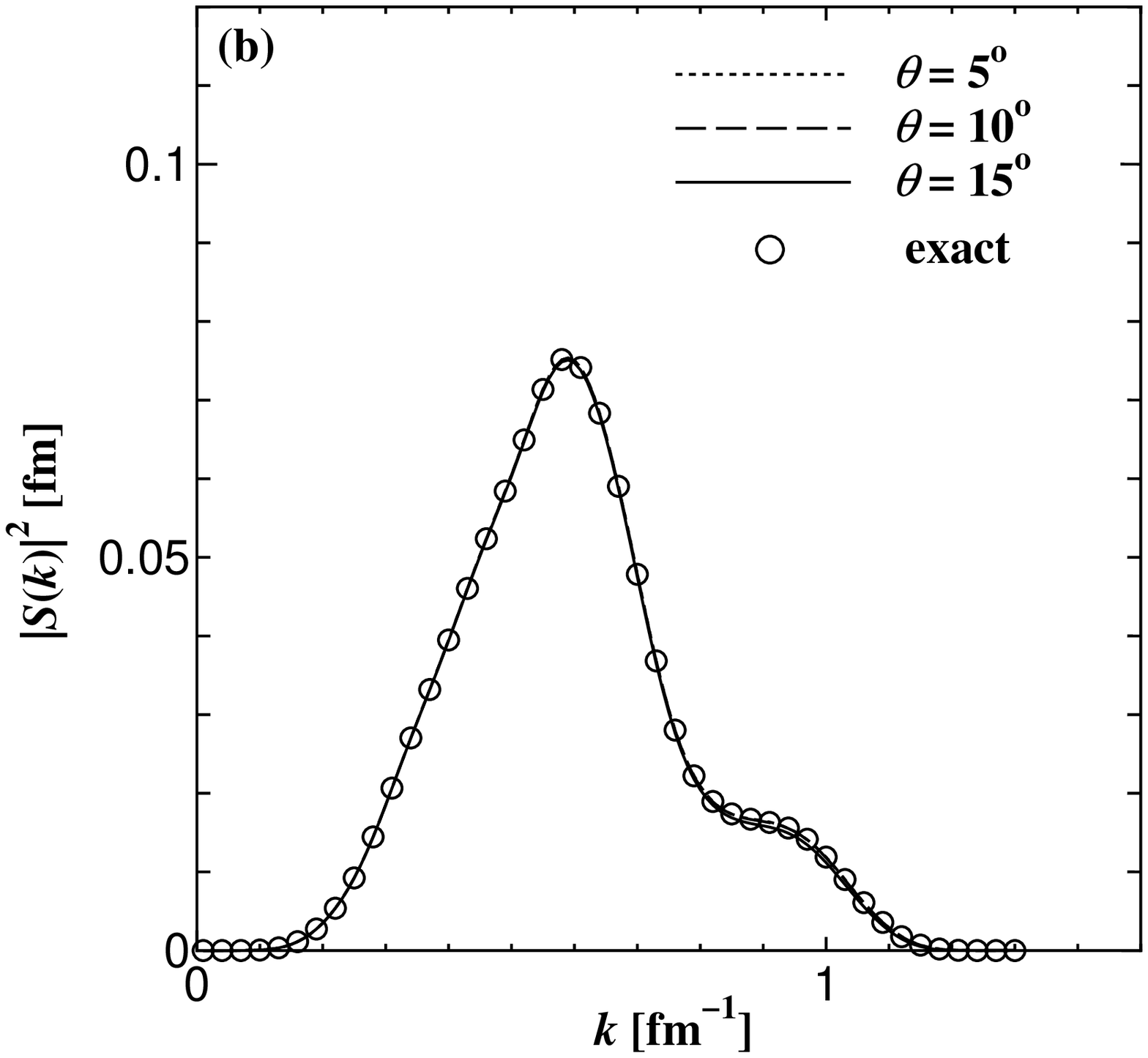}
 \includegraphics[clip,width=0.4\textwidth]{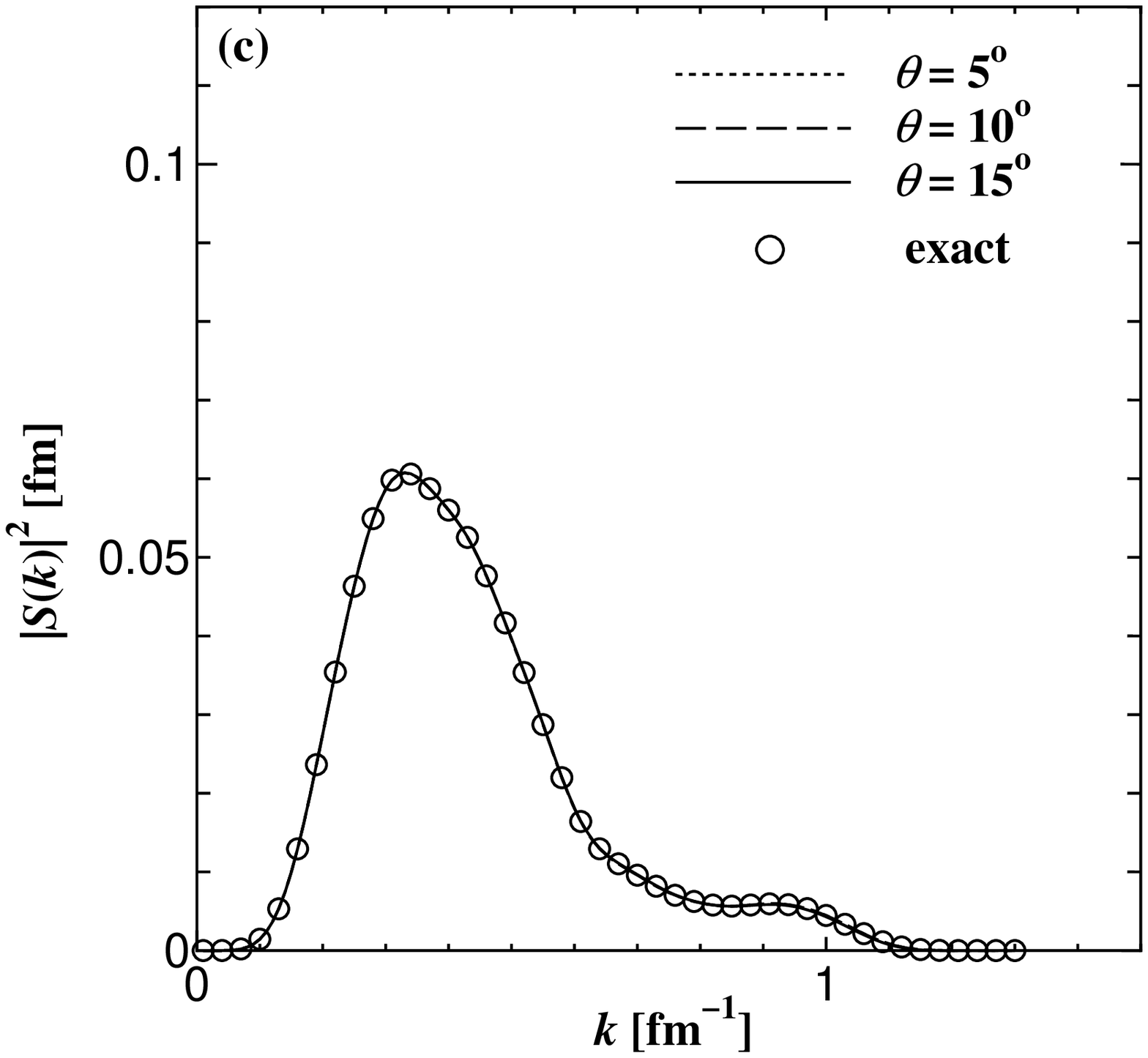}
 \includegraphics[clip,width=0.4\textwidth]{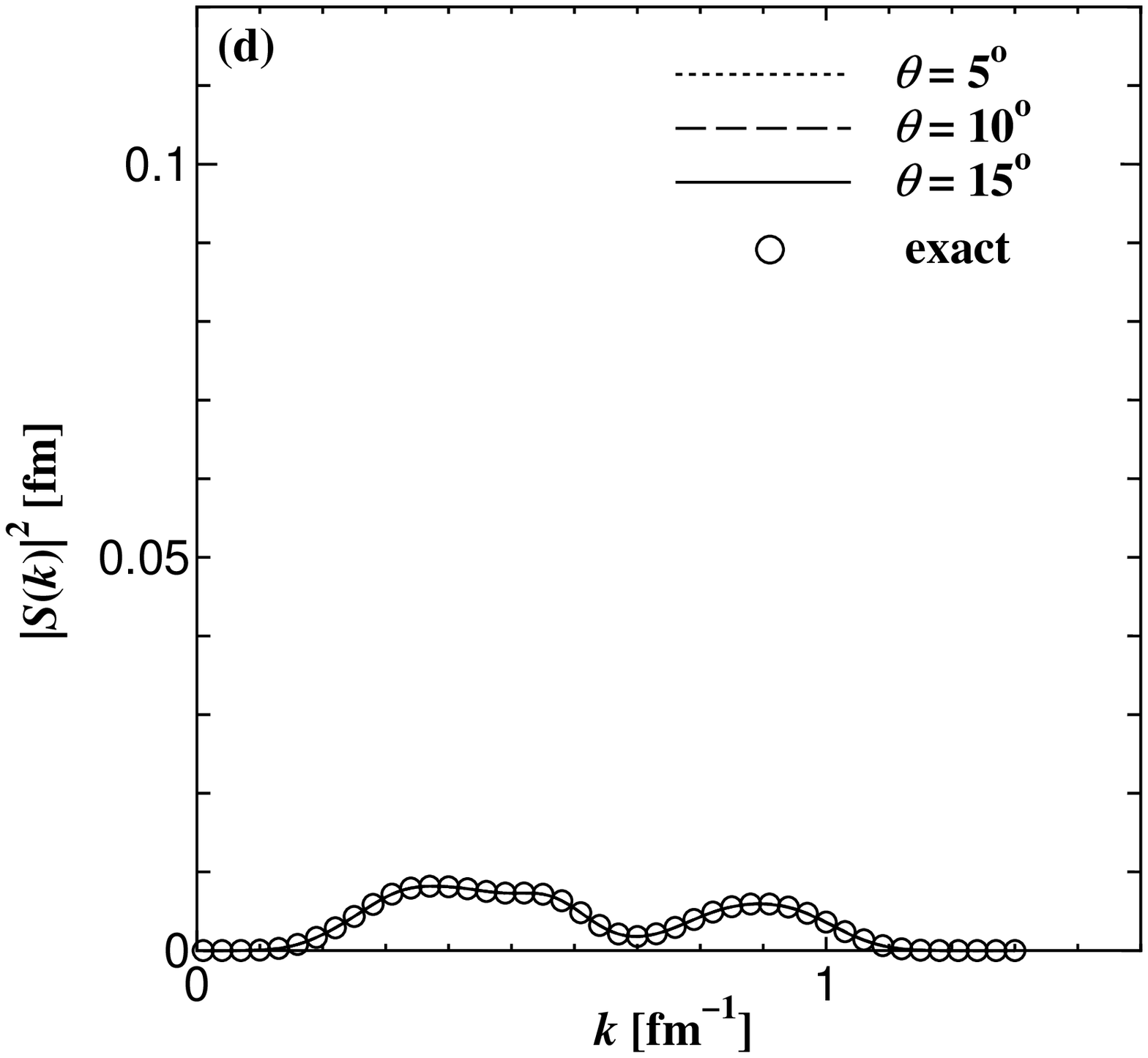}
 \caption{The squared moduli of breakup $S$-matrix elements as a
 function of $k$ at the grazing total angular momentum $J=17$ for
 $d$+$^{58}$Ni scattering at 80 MeV. Panels (a), (b), (c), and (d)
 correspond to $(\ell,L)=(0,17)$, (2,15), (2,17), and (2,19),
 respectively.}
 \label{smat-d}
\end{center}
\end{figure}%
\subsection{$^{58}$Ni($d$,$pn$) reaction at 80 MeV}

In the analysis of the $^{58}$Ni($d$,$pn$) reaction, we adopt the same
model Hamiltonian as in the previous paper.~\cite{Matsumoto} \
As for the breakup states of $d$ we take $\ell=0$ and 2 states
and truncated them at $k=1.5$ fm$^{-1}$. The range parameters
of the complex-range Gaussian basis functions
are $(2N=40,a_1=1.0~\mbox{fm},a_N=30.0~\mbox{fm})$.
In the calculation of the complex scaled states, we take
$(N=30,a_1=1.0~\mbox{fm},a_N=30.0~\mbox{fm})$ as the parameter set
of the real-range Gaussian basis functions.

Figure \ref{smat-d} shows the squared moduli of breakup $S$-matrix
elements as a function of $k$ at the grazing total angular momentum
$J=17$; panels (a), (b), (c) and (d) correspond to ($\ell,L$)=(0,17),
(2,15), (2,17) and (2,19), respectively. Here $L$ is the orbital
angular momentum regarding ${\bf R}$. The open circles are the results
calculated with the exact smoothing factors that are obtained by solving
the two-body Schr\"odinger equation \eqref{eq:Sch-B} numerically.
The dotted, dashed, and solid lines correspond
to results of the new smoothing method with CSM at the scaling angle
$\theta=5^\circ$, 10$^\circ$, and 15$^\circ$, respectively. One can see
that the results converge to the exact solution as the scaling
angle $\theta$ increases.

\begin{figure}[tb]
\begin{center}
 \includegraphics[clip,width=0.4\textwidth]{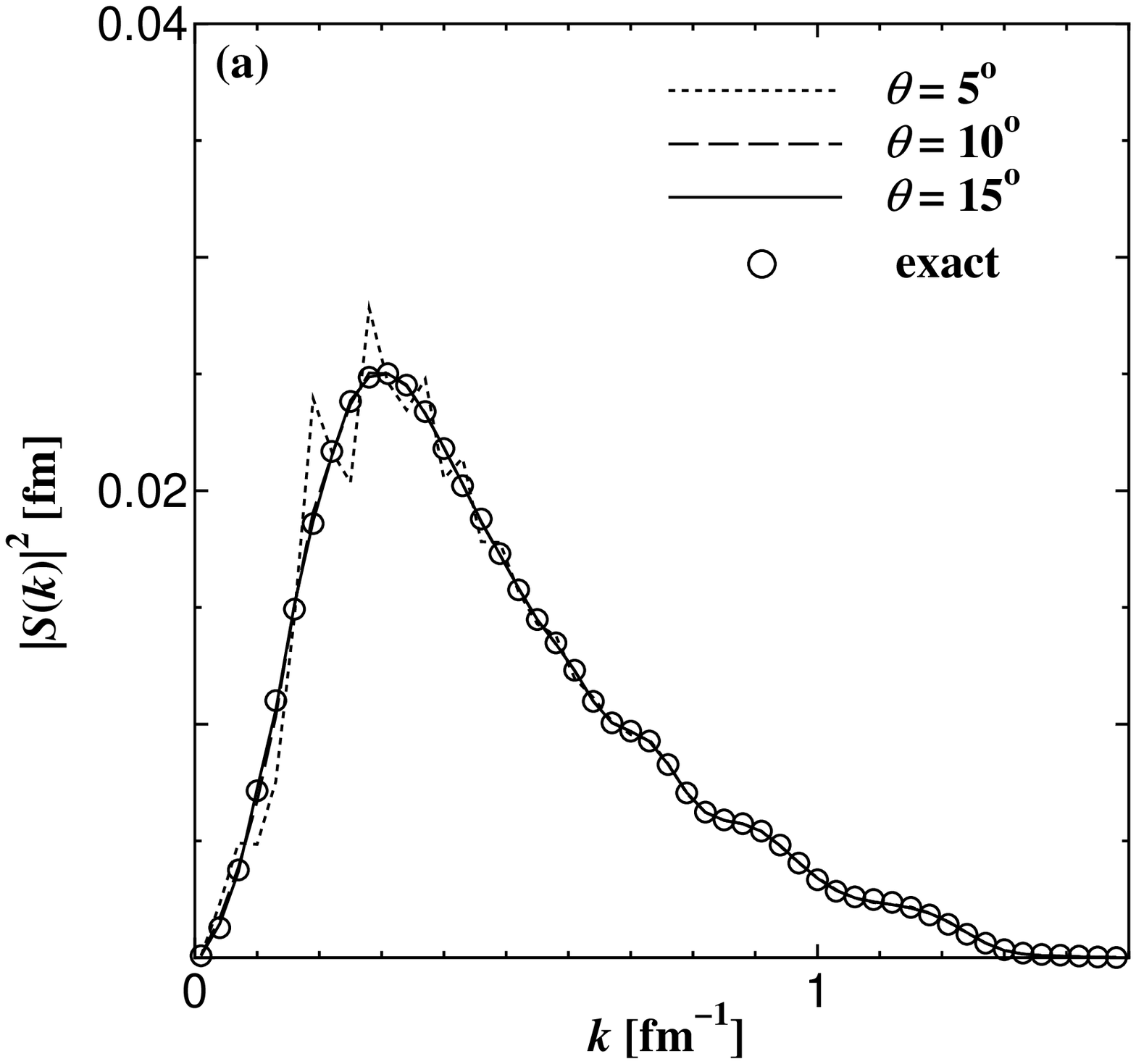}
 \includegraphics[clip,width=0.4\textwidth]{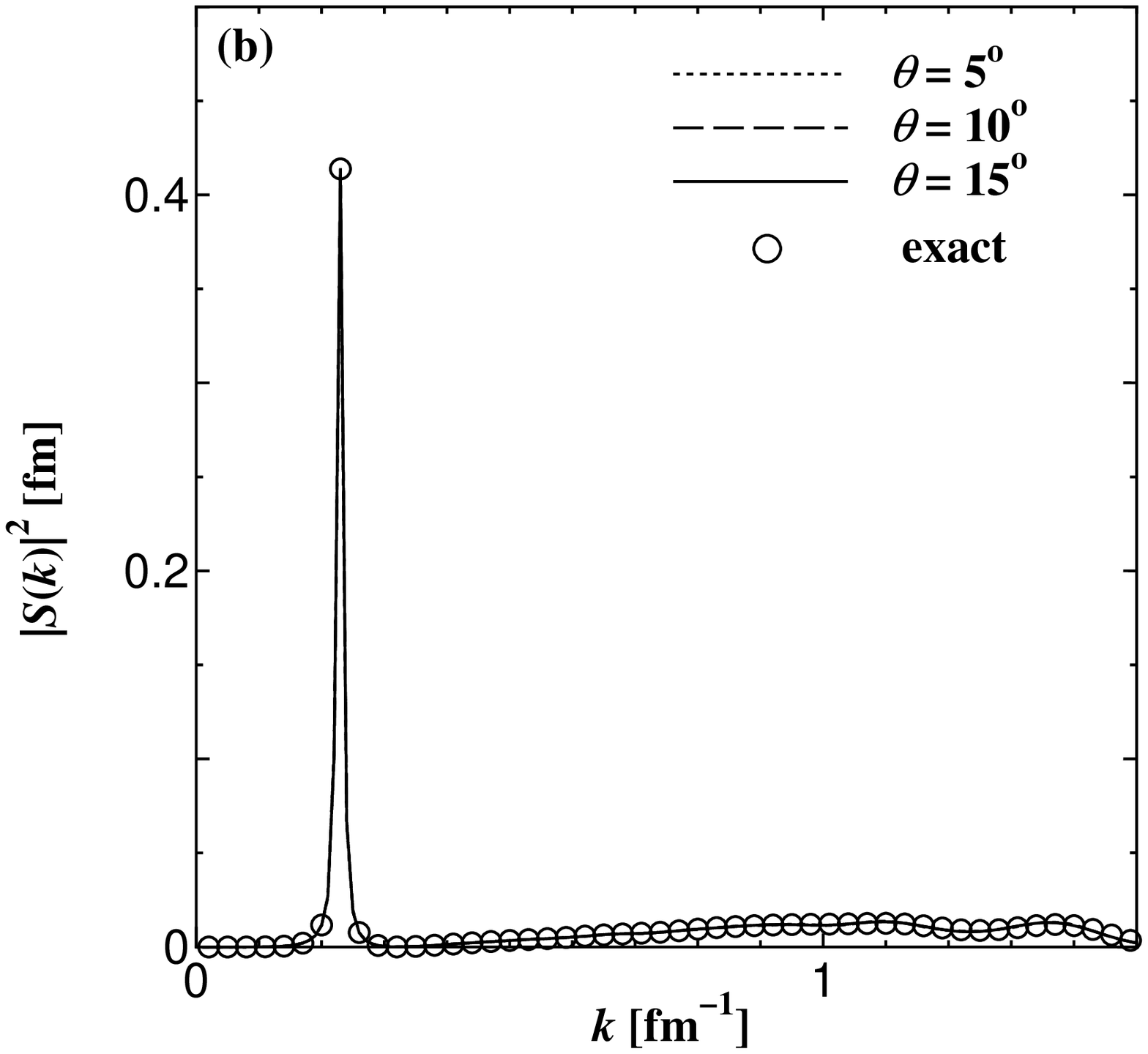}
 \includegraphics[clip,width=0.4\textwidth]{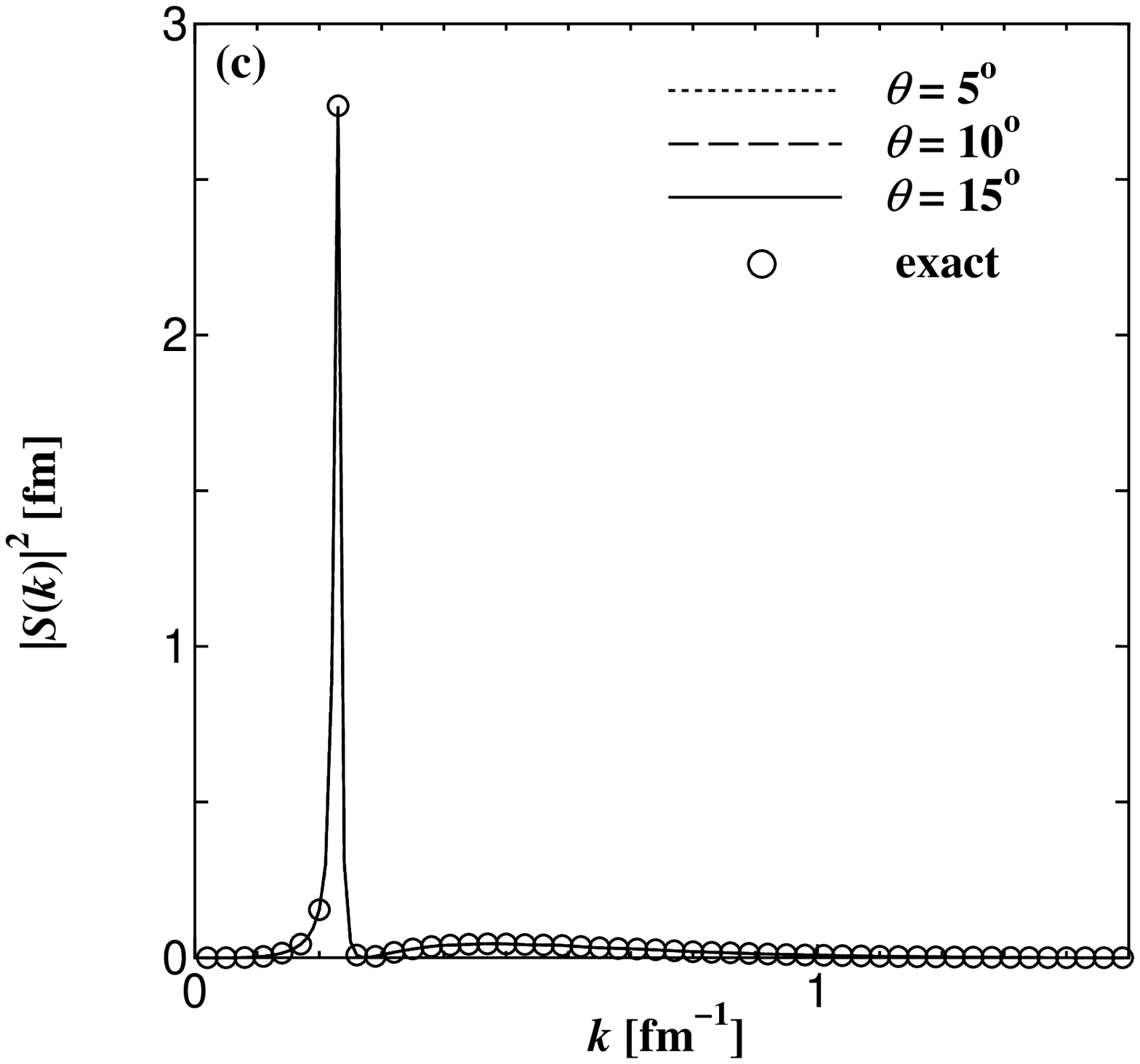}
 \includegraphics[clip,width=0.4\textwidth]{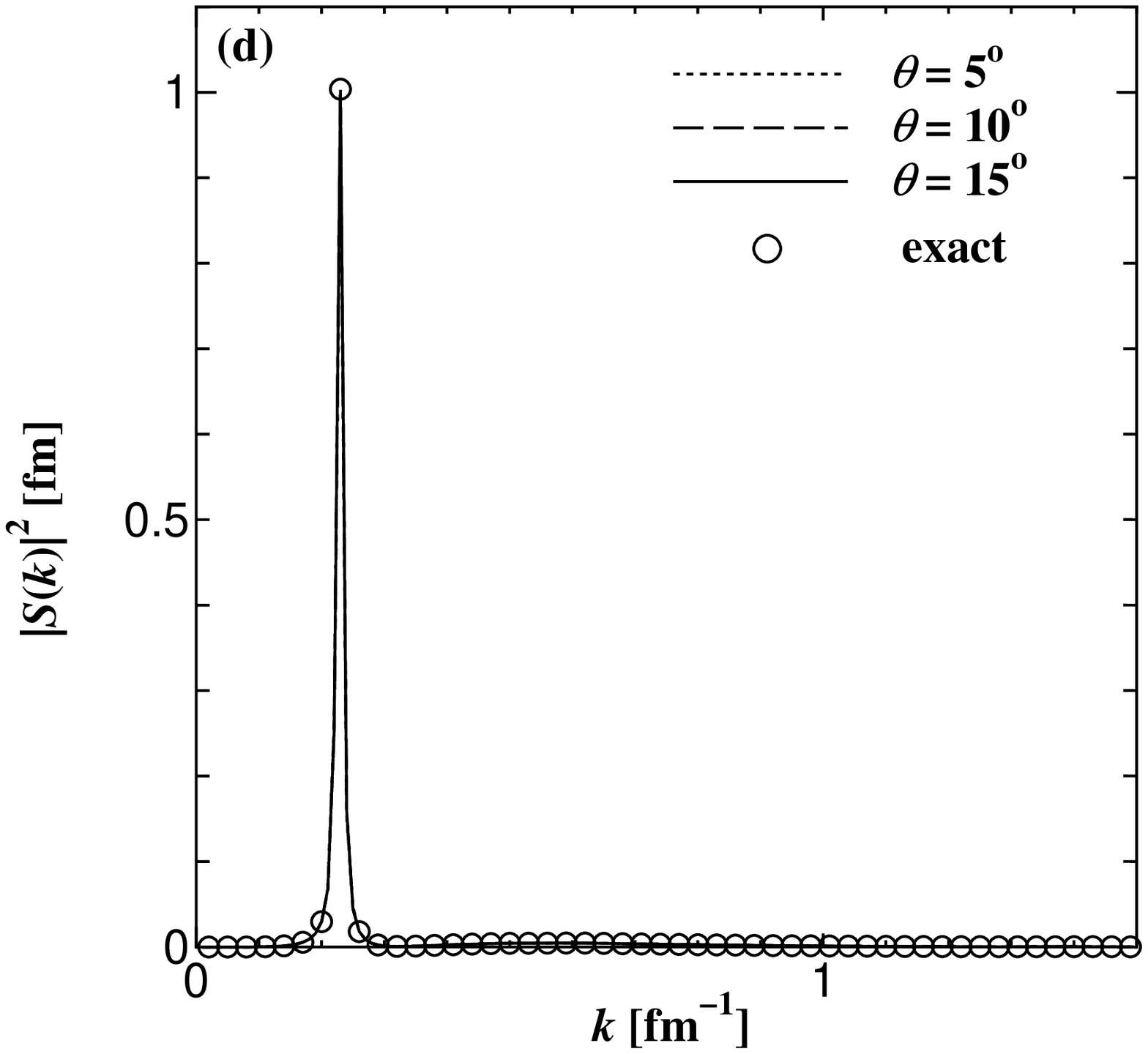}
 \caption{The same as in fig.~\ref{smat-d} but for
 $^6$He+$^{12}$C scattering at 229.8 MeV. The corresponding grazing
 angular momentum is 25. Panels (a), (b), (c), and (d) represent
$(\ell,L)=(0,25)$, (2,23), (2,25), and (2,27), respectively.}
 \label{smat-6He}
\end{center}
\end{figure}%
\subsection{$^{12}$C($^6$He,$^4$He$^2n$) reaction at 229.8 MeV}

We take $\ell=0$ and 2 breakup states and truncate them at $k=$1.5 fm$^{-1}$.
We use the following interaction $V_{^2n\alpha}$ between $^2n$ and $^4$He:
\begin{eqnarray}
V_{^2{\rm n}\alpha}(r)&=&V_0\exp\left[-(r/\beta)^2\right] ,
\label{2n-a_int}
\end{eqnarray}
where the depth $V_0$ and the range $\beta$ are determined so as to
reproduce the ground-state and resonance energies of $^6$He.
As for the interactions between $^{12}$C and each constituent
of $^6$He, i.e. $^6$He and $^2n$,  we take
the optical potentials of $d$+$^{12}$C~\cite{d-C}
and $^4$He+$^{12}$C.~\cite{a-C}
The parameters of the complex-range Gaussian basis functions are
($2N=40$, $a_1=1.0$ fm, $a_N=30$ fm),
and those of the real-range Gaussian basis functions are
$(N=30,a_1=1.0~\mbox{fm},a_N=30.0~\mbox{fm})$.

Figure \ref{smat-6He} represents the squared moduli of
the breakup $S$-matrix elements
at the grazing total angular momentum $J=25$.
Panels (a), (b), (c), and (d) correspond
to $(\ell,L)=(0,25)$, (2,23), (2,25), and (2,27), respectively.
The peaks shown in
panels (b), (c), and (d) represent the $2^+$ resonance of
$^6$He. The breakup $S$-matrix elements calculated with the simple
smoothing formula Eq.~\eqref{eq-for-smoothing} tend
to the exact ones (the open circles) as $\theta$ increases.
One can conclude from Fig.~\ref{smat-6He} that the
smoothing formula is valid also for the resonance part.

\section{Summary}
\label{sec:summary}

In this Letter, we present a practical formula for smoothing discrete
breakup $S$-matrix elements calculated by CDCC.
This smoothing procedure based on CSM can easily be performed by diagonalizing
$H_{\rm B}$ and the complex scaled one $H_{\rm B}^{\theta}$
with the $L^2$-type bases such as the Gaussian basis function
and inserting the eigenenergies and eigenstates into the formula.
The validity of the formula is tested
for two kinds of three-body breakup reactions,
$^{58}$Ni($d$, $p n$) at 80~MeV and
$^{12}$C($^6$He, $^4$He$^2n$) at 229.8~MeV.
For both the cases, the breakup S-matrix elements smoothed with
the formula tend to the exact ones as the complex-scaling angle $\theta$
increases. Thus, the formula is accurate and practical.
In a forthcoming paper, we will investigate the practicability of
this formula for four-body breakup processes of three-body projectiles such
as $^6$He and apply this method for analyzing
the experimental data~\cite{Aumann,Wang-MSU,Chulkov}
on the $^6$He breakup reactions
at lower and intermediate incident energies.

\section*{Acknowledgements}

This work has been supported in part by JSPS Research Fellowships
for Young Scientists and the Grants-in-Aid for Scientific Research of
Monbukagakusyou of Japan.
The numerical calculations of this work were performed on the computing
system in Research Institute for Information Technology of Kyushu University.


\end{document}